\documentstyle[aps,prl,twocolumn,psfig]{revtex}
\begin{document}
\title{A simple deterministic self-organized critical system}
\author{Maria de Sousa Vieira\cite{email}}
\address{Department of Biochemistry and Biophysics, University of California,
San Francisco, California 94143-0448.}
\maketitle
\begin{abstract}
We introduce a new continuous cellular automaton 
that  presents self-organized criticality. 
It is one-dimensional, totally deterministic, without any 
embedded randomness, not   
even in the initial conditions.  
This system is 
in the same universality class as the Oslo rice pile, 
boundary driven  interface depinning  
and the train model for earthquakes.  
Although the system is chaotic, in the thermodynamic limit 
chaos occurs only in a microscopic level. 
\end{abstract}
\narrowtext
In 1987, Bak, Tang and  
Wiesenfeld 
showed that    
fractal behavior, that is, power-law distributions,  
can be observed in simple dissipative 
systems with 
many degrees of freedom without  fine tuning of parameters\cite{soc}. 
They called this phenomenon self-organized criticality (SOC). 
Until then, the studies of fractal structures  
were basically related to equilibrium systems where fractality 
appears only at special parameter values where a
phase transition takes place. 

Since the pioneering work of Bak {\sl et. al}, 
an enormous amount of numerical, theoretical and experimental studies 
have been done in systems that present SOC.
One of the most interesting experimental studies demonstrating the 
existence of SOC in Nature was done in a 
quasi-one-dimensional pile of rice 
by Frette {\sl et. al}\cite{frette}. 
They found that 
the occurrence of SOC depends on the shape of the rice. 
Only with sufficient 
elongated grains, avalanches with a power-law distribution 
occurred. If the rice had little asymmetry, a distribution described by  
a stretched exponential was seen. Christensen {\sl et. al}\cite{oslo} 
introduced 
a model for the rice pile experiment in which the local 
critical slope varies 
randomly between 1 and 2. They found that their model, 
known as the  Oslo rice pile model, reproduced 
well the experimental results.  

A good understanding of the Oslo system was achieved by  
Paczuski and Boettcher\cite{boettcher}. They showed that it could 
be mapped exactly to a model for interface depinning where the interface 
is slowly pulled at one end through a medium with quenched random 
pinning forces. They found that the height of the interface 
maps to the number of toppling events in the rice pile model. 
The critical  exponents of the two models were identical (within 
the error bars), showing 
that they were in the same universality class.  
Paczuski and Boettcher also conjectured 
that the train model for earthquakes, which was introduced by Burridge and 
Knopoff\cite{bk}, and studied in detail in \cite{train}, is also 
in that same universality class. 
The train model is the only 
model that we know (besides the one we introduce here) 
that presents SOC and has no kind 
of embedded randomness. However, it is governed by
coupled ordinary equations (ODE's), what makes its study 
very time consuming. 

A way of making a system governed by ODE's  
more amenable to computer 
simulations is to discretize it in time.   
This was done by Olami, Feder and Christensen (OFC)\cite{ofc} who 
introduced a 
continuous cellular automaton (CCA) to  
study the two-dimensional version of another Burridge and Knopoff 
model for earthquakes\cite{bk}. 
[A continuous cellular automaton in SOC is known in chaos theory 
as coupled lattice maps. These systems are characterized as 
having space and time variables defined in the domain of real and 
integer numbers, respectively.]    
In the OFC model,   
SOC is seen only 
in systems that have a geometry with dimensionality 
of at least two. That system is a variation of 
a model introduced by Nakanishi\cite{naka}, which 
has a one-dimensional geometry. 
However, the model introduced by Nakanishi {\sl does not}  
present SOC, since the power-law
distribution it presents has an upper cutoff that is unrelated to the 
system size. 

Here we introduce a new self-organized critical system, 
that is governed by a CCA (that is, the space is continuous and 
the time is discrete). 
It  is one-dimensional and has no embedded randomness, not 
even in the initial conditions. 
We will show that our model belongs 
to the same universality class as the Oslo rice pile, 
boundary driven interface 
depinning and the train model. 
The importance of our results comes from the fact that 
we show that it is possible to map stochastic SOC systems  
to simple, discrete, chaotic systems, in which 
no randomness exists.  Such an 
equivalence of a deterministic model with no embedded 
randomness which is chaotic with a stochastic model also 
occurs between the deterministic Kuramoto-Shivashinsky\cite{kuramoto}
equation and the Langevin equation proposed by Kardar, Parisi and 
Zhang\cite{kardar}. In our opinion, the train model 
governed by ODE's already achieved this\cite{train}. However, because 
its simulation is very time consuming, it will probably be impossible 
to find such equivalences for higher dimensional (two, three, etc.) systems, 
unless the system is discretized in time, as we do here. In fact, we 
are unaware of any studies on  train-like systems with dimensionality 
higher than one.   
To the best of our knowledge, our model is the only SOC system introduced 
so far   
that is one-dimensional, totally 
deterministic and with discrete time. 

Another important result of this paper concerns to the fact that  
we show that although chaos exists in the 
model, it decreases as the system size increases, and in 
thermodynamic limit it exists only in a microscopic level.  
Consequently, our results 
indicate that the fractal structures seen in 
Nature and supposed associated with SOC, could in principle 
result only from nonlinearities in those systems, without 
any need for the presence of random irregularities. 
Such fundamental questions are also found in 
equilibrium statistical mechanics, where it is 
unknown if probability theory is only a tool to 
describe phenomena that in principle could have been   
originated solely from microscopic chaos\cite{zav}.   
    
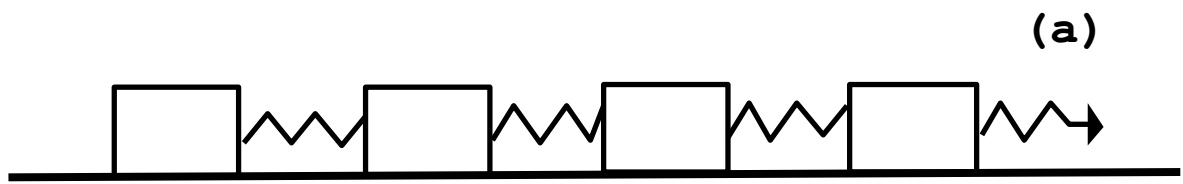
\begin{figure}
\caption[f1]{
(a) The train model. (b) The function $\phi (x)$.
}
\label{f1}
\end{figure}

The train model is shown in 
Fig.~\ref{f1}(a). It 
consists of a chain  of blocks connect by harmonic springs. The blocks 
are on a rough surface with friction, and  
the first block is pulled slowly  
with a constant velocity by a driver. 
The dynamics of the model is 
as follows: suppose that all the blocks are initially at rest.
As the driver pulls the first block, it remains 
stuck until the elastic force overcomes the static friction. 
When this occurs, the first block will move a little and will be  
stopped again by friction. Such small events (or earthquakes) will continue, 
and will increase the elastic force on the second block. 
There will be a moment when the elastic force on the second 
block will overcome the friction force, and then we will 
see an event involving two blocks. This      
kind of dynamics will continue with events involving three, four,   
or all the blocks in the system. 

The elastic force in block $i$ is given by 
$f_i=x_{i-1}-2x_i+x_{i+1}$, where $x_{i}$ is the displacement of 
block $i$ with respect to its equilibrium position (without 
losing generality, the spring constant can be taken as equal 
to 1\cite{train}).
The boundary conditions are $x_0=0$ and $x_{L+1}=x_{L}$. 
After an earthquake, in which block $i$ was displaced by 
$\Delta x$, the elastic forces in block $i$ and in its 
nearest neighbors will be $f'_i=f_i-\Delta f$, and 
$f'_{i\pm 1}=f_{i \pm 1} +\Delta f/2$, respectively, where 
$\Delta f = 2\Delta x$. 
Thus,
the force that is relaxed in block $i$ is redistributed equally to its 
near nearest neighbors, implying conservation of elastic forces.  
This is embedded into the geometry of the system. However, 
the model does 
dissipate energy through friction between the rough 
surface and the blocks. Consequently, 
the model is conservative with respect to elastic forces, 
and dissipative with respect to energy.  
This  is one of the main distinctions between the 
train model and the other Burridge and Knopoff  earthquake model studied 
by Nakanishi and OFC, in which neither the energy nor 
the forces are conserved.  

In the discretized version of the train model that we introduce here, 
each block $i$ is characterized by a variable $f_i$, which we 
will call force, with  
$i=1,..., L$, and $L$ being  the number of blocks in the system. 
The boundary conditions are the same as the ones in the train model, 
which are given above. 
The dynamical evolution of the system is determined by 
the following algorithm:

(1) Start the system by defining initial values for the variables 
$f_i$, which can be the same for all the blocks,  
so the they are 
below a chosen, fixed,  threshold $f_{th}$.     

(2) Update the force in the first block by incrementing its force 
to the threshold value plus a fixed small value $\delta f$, i.e., 
$f_1=f_{th}+\delta f$ (an event is going to be triggered).   

(3) Check the forces in each block. If a block $i$ has $f_i \ge f_{th}$, 
update  $f_i$ according to $f'_i=\phi(f_i-f_{th})$, where 
$\phi $ is a given nonlinear function that has a parameter $a$. 
Increase the forces in its two nearest neighboring blocks according 
to  $f'_{i \pm 1}=f_{i\pm 1}+\Delta f/2$, where 
$\Delta f=f_i-f_i'$.   

(4) If $f'_i < f_{th}$ for all the blocks, go to step (2) (the event 
has finished). Otherwise, go 
to step (3) (the event is still evolving).    

One can use either parallel or sequential update in the evolution of 
the system. We have verified 
that the critical exponents of the model do not depend on the type 
of update chosen.   
The systems is  governed by $L$ variables and has two 
parameters, $a$ and $\delta f$, since 
without losing generality  we can take $f_{th}=1$.   
The force in our model is supposed to mimic the combination 
of two forces in the train model, that is, the elastic and 
the friction forces. The elastic force is periodic, whereas 
the friction force in simulations is generally assumed 
to decrease with 
the velocity of the block. We have found numerically that 
$\phi $ mimics the combination of 
these two force when it is a periodic function, since only in this 
way the system presents SOC behavior. So, the periodicity of 
the elastic force dominates over the form of the friction force.  
The models introduced by   
Nakanishi and OFC assume that 
$\phi $ is a strictly nonincreasing function.  
We have found that if we use a strictly nonincreasing 
function for $\phi $, such as the one used in\cite{naka},  
we observe in our model the presence of 
stretched exponentials, instead of power-laws\cite{future}.  
It is worth noting that in this situation our model reproduces the 
results of the distributions found with 
nearly round rice 
in \cite{frette}, which were also governed by 
stretched exponentials.

The functional form we chose for  $\phi$ is 
shown in Fig.~\ref{f1}(b), 
which is given by, 
$\phi (x) = 1-a[x]$  
where $[x]$ denotes $x \ {\rm modulo} \ 1/a$, 
that is, a sawtooth function. 
However, we have tested several other periodic functions,  
and found that the SOC behavior we show here remains, that is,  
the results are robust, the essential ingredient being periodicity 
(not necessary a perfect one) 
for $\phi $.   

In Fig.~\ref{f2} we show the distribution of events involving 
$s$ update steps, which is the size of the event, using parallel
update.  
The events that involve all the 
blocks of the chain have been excluded from our analysis, since  
they do not belong to the same distribution, as 
expected. 
Before we start to compute the statistics, we 
wait until the last block has moved, in order to neglect transient 
effects.  
In (a) we show three different cases, with $L=512$,  in which we have varied 
one parameter at a time. 
The solid line refers to $a=4$ and $\delta f=0.1$, the dashed line is for 
$a=4$ and $\delta f=0.01$, and the short-dashed line refers to $a=2.5$ and 
$\delta f=0.01$. 
We see that the small events have their own distribution, like in 
the Oslo model for rice piles\cite{oslo}. 
A careful analysis has shown that these small events have an exponential 
distribution\cite{future}. 
As $s$ is increased 
the distribution becomes a power-law, which has a cutoff related to  
finite size effects, only.
We find that the slope of 
the power law is independent of $a$ and $\delta f$.  
However, the crossover point $s^*$ of the exponential  
behavior 
to the power law one depends on $a$, but not 
on $\delta f$. The frequency of the small events      
is inversely proportional to $\delta f$. 

In Fig.~\ref{f2}(b) we show simulations 
keeping $a$ and $\delta f$ fixed, with 
$a=4$ and $\delta f=0.1$, and 
varying $L$. We see that increasing $L$, the range of the power-law 
increases. To fit the data to a single curve, we try the finite 
size scaling ansatz   
$P(s,L)=s^{-\tau}G(s/L^D)$,    
where $D$ and $\tau $ are the basic exponents of the model\cite{bakp}, 
defining its universality class. 
$D$ and $\tau $ are called the dimension and the  
size distribution exponents, respectively. 
In our model we find that $<s> \sim L$ which results in
$\tau=2-1/D$. 
The best 
fit for $P(s,L)$ is found for $\tau \approx 1.54$ and $D \approx 2.20$. 
The data 
collapse for these values of the exponents is shown in  
Fig.~\ref{f2}(c). Within the error bars, 
these exponents are the same ones of the Oslo rice pile\cite{frette,boettcher},
driven boundary interface 
depinning\cite{boettcher} and the train model for 
earthquake\cite{boettcher,future}. 
Consequently, all these  models, including the one we introduce here, 
are in the same universality class.

In Fig.~\ref{f3}(a) we show the frequency of 
the events $P(T)$ as a function of the its time duration $T$, 
for different system sizes. The parameters are the same 
as in Fig.~\ref{f2}(b).  Again, we see a power-law distribution, except 
for the smallest events. A data collapse for the 
function $P(T,L)=T^{-y}f_T(T/L^{\sigma})$ is shown in Fig.~\ref{f3}(b), 
with $y=1.84$ and $\sigma=1.40$. These are the same 
exponents found in the Oslo rice pile.  From 
conservation of probability it follows that $\sigma(y-1)=D(\tau-1)$ in 
good agreement with our results.   
The results shown in Fig. 3 are, again, for parallel update, since 
in sequential  update the time duration and event size coincide resulting 
in $y=\tau$ and $\sigma=D$.   

Using the method introduced by Benettin {\sl et al.}\cite{liap}, 
we have calculated the largest Liapunov exponent, $\lambda_1$,   
and 
the second largest Liapunov exponent, $\lambda_2$, of the system.  
If $\lambda_1$ is greater than zero, 
it implies that 
the system has a strong sensitivity to the initial conditions, and 
by definition, it is called chaotic. To study the Liapunov 
exponents we have chosen sequential update. The 
reason for this is that the calculation of the Liapunov 
exponent assumes, by its own definition, that small changes 
happen in the system in the time unit, and this is more consistent with 
sequential rather than with parallel  update.  
In Fig.~\ref{f4}(a) and Fig.~\ref{f4}(b)   
we show $\lambda_1$ and $\lambda_2$ 
as a 
function of $a$ for $\delta f=0.1$, 
and as a function of $\delta f$ for $a=4$, respectively.
In both cases  $L=64$. We have 
found that for $a \le 1$ the system is in continuous motion, 
and therefore being impossible to define earthquakes. 
Consequently, SOC is only seen when $a>1$. 
We see that the Liapunov exponents 
increase as $a$ or $\delta f$ increases, with the 
other parameters kept fixed. 
Fig.~\ref{f4}(c) shows the largest Liapunov exponent as 
a function of 
the system size for $a=4$ and $\delta f=0.1$ (solid), 
$a=4$ and $\delta f=0.01$ (dashed) and $a=2.5$ and $\delta f=0.01$ 
(short-dashed). 
We observe that $\lambda _1$ is approximately constant for 
small $L$ and decreases nearly as power-law when $L$ is greater 
than a certain value. The value where this bending occurs 
seems to be sensitive to both $\delta f$ and $a$.
Since $\lambda _1 \to 0$   
in the thermodynamic limit
($L \to \infty$) we conclude that  chaos exists only in 
a microscopic level, and that any time or space scales in the system 
are negligible. 
We have studied the system using slower time  
scales, such as measuring time by the updates in the 
first block. Still we find that  $\lambda _1 \to 0$ when $L \to \infty$.
In the train model governed by ODE's and pulled with  
a constant finite velocity  
we have 
found that the largest Liapunov exponent tends to a constant as 
the system size increases\cite{liapt}. 
However, our new unpublished results shows that the Liapunov
exponents in that system  
start to decrease for system sizes  greater than a given 
value, as it happens in the system we introduce here.     


\begin{figure}
\caption[f2]{
Probability distribution $P(s)$ of the number of toppling events $s$. 
(a) solid: $L=512$, $a=4$, $\delta f=0.1$; dashed:
$L=512$, $a=4$, $\delta f=0.01$; short-dashed: 
$L=512$, $a=2.5$, $\delta f=0.01$.  
(b) $L=32, 64, 128, 256, 512$ with  $a=4$ and $\delta f=0.1$. (c) 
Data collapse  of the cases shown in (b) with $\tau = 1.54$ and 
$D=2.20$. 
}
\label{f2}
\end{figure}

\begin{figure}
\caption[f3]{
(a) Probability distribution $P(T)$ of the number of toppling events $T$ for 
$L=32, 64, 128, 256, 512$ with  $a=4$ and $\delta f=0.1$. (b)
Data collapse of the cases shown in (a) using  
$P(T)=T^{-y}f_T(T/L^{\sigma})$ 
with $y = 1.84$ and
$\sigma =1.40$. 
}
\label{f3}
\end{figure}

\begin{figure}
\caption[f4]{
(a) The largest (solid) and the second largest (dashed) Liapunov exponents  
as a function of $a$ for $\delta f=0.1$, and  (b) 
as a function of $\delta f$ for $a=4$, with $L=64$ in both cases.  
(c) The largest Liapunov exponent as a function of $L$ for 
$a=4$ and  $\delta f=0.01$ (dashed),  
$a=2.5$ and  $\delta f=0.01$ (short-dashed).  
The error bars associated with 
this calculation have approximately the size of the plus sign shown on the 
curves. 
}
\label{f4}
\end{figure}


\begin{references}
\bibitem[*] {email} Electronic address: mariav@msg.ucsf.edu. 

\bibitem {soc} P. Bak, C. Tang, and K. Wiesenfeld,  Phys. Rev. Lett.
{\bf 59}, 381 (1987);  P. Bak {\sl et. al}, 
Phys. Rev. A {\bf 38}, 364 (1988).

\bibitem{frette} V. Frette {\sl et. al}, Nature (London) {\bf 379}, 49 (1996).

\bibitem{oslo} K. Christensen {\sl et. al}, Phys. Rev. Lett. {\bf 77}, 107 (1996).

\bibitem{boettcher} M. Paczuski and S. Boettcher, Phys. Rev. Lett. 
{\bf 77}, 111 (1996).

\bibitem {bk} R. Burridge and L. Knopoff, Bull. Seismol. Soc. Am. 
{\bf 57}, 
341 (1967).

\bibitem{train} M. de Sousa Vieira,  Phys. Rev. A {\bf 46}, 6288 (1992). 

\bibitem{ofc} Z. Olami {\sl et. al}, Phys. Rev. Lett. {\bf 8}, 1244 (1992).

\bibitem {naka} H. Nakanishi, Phys. Rev. A {\bf 43}, 6613 (1991). 


\bibitem{future} M. de Sousa Vieira, {\sl to be published}. 

\bibitem{kuramoto} V. Yakhot, Phys. Rev. A {\bf 24}, 642 (1981).

\bibitem{kardar}  M. Kardar, G. Parisi, and Y.-C Zhang, Phys. Rev. Lett. 
{\bf  56} 889 (1986).   

\bibitem{zav} G. M. Zaslavsky, Phys. Today {\bf 52}, 39 (1999).  


\bibitem{bakp} M. Paczuski, S. Maslov and P. Bak, Phys. Rev. E {\bf 53}, 
414 (1996).

\bibitem {liap} G. Benettin {\sl et al.}, Phys. 
Rev. A {\bf 14}, 2338 (1976). 

\bibitem{liapt}   M. de Sousa Vieira {\sl et. al.}, Phys. Rev. E {\bf 53}, 
1441 (1996).    

\end{references}
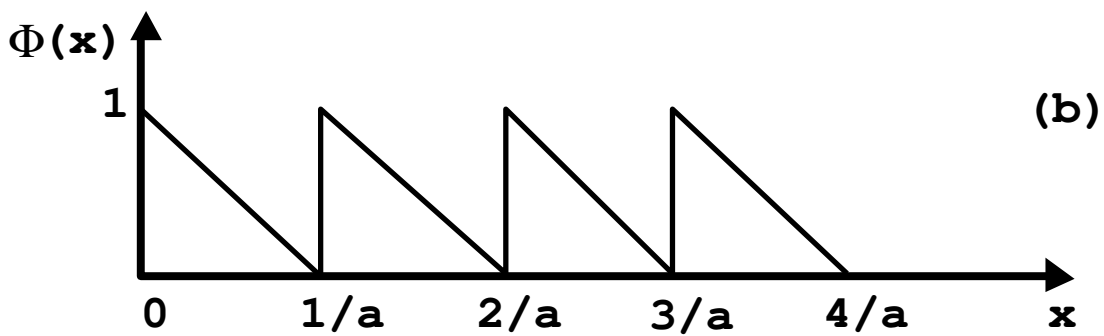
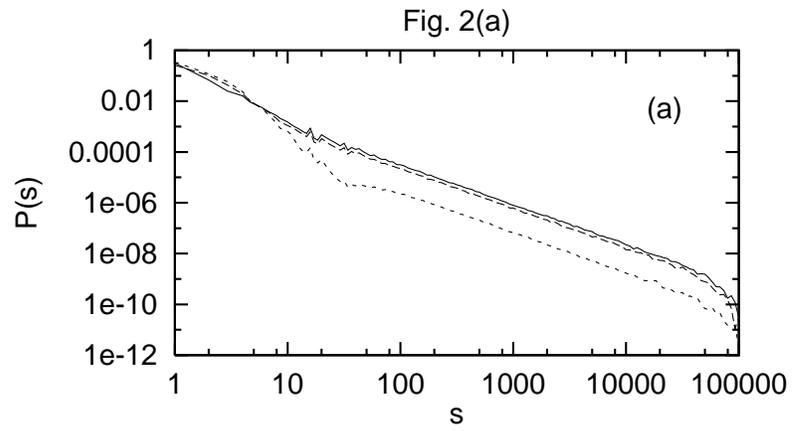
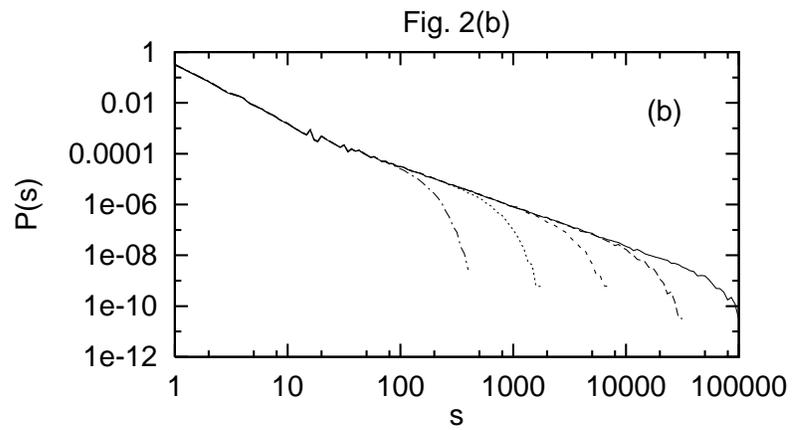
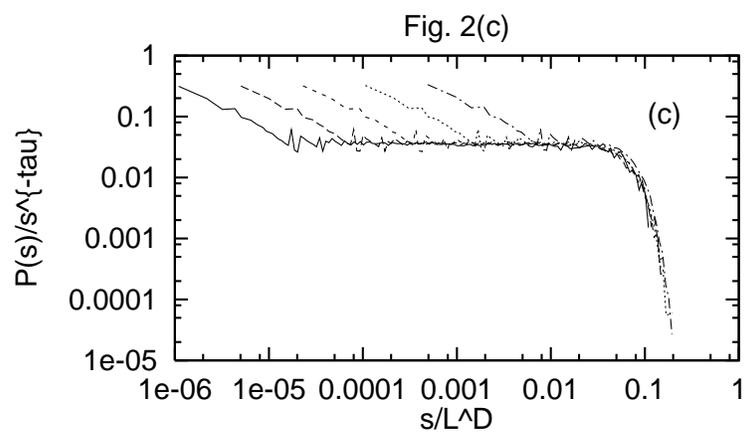
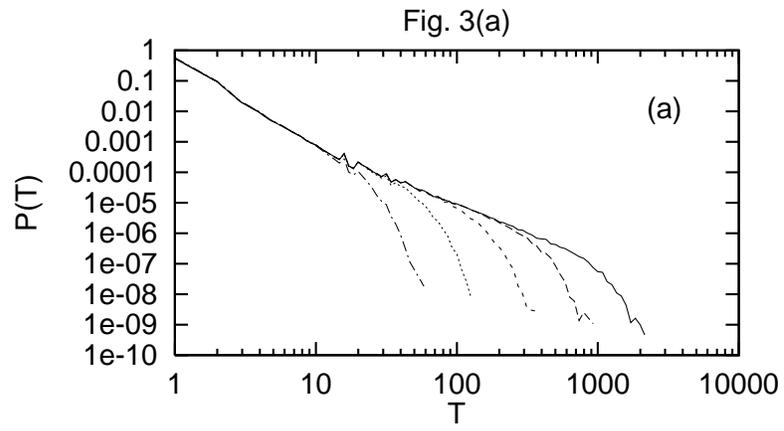
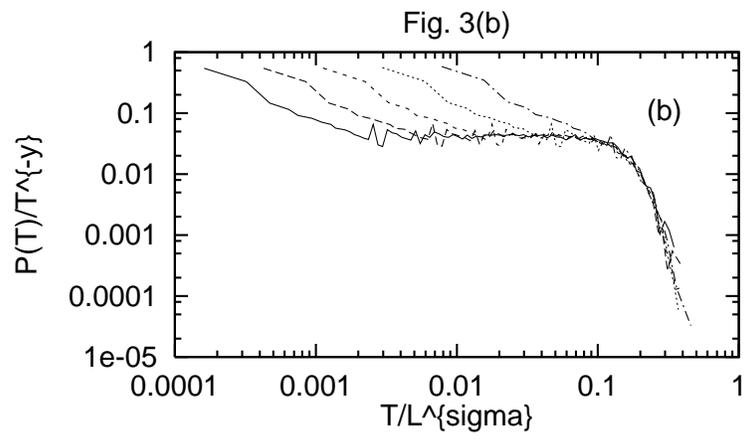
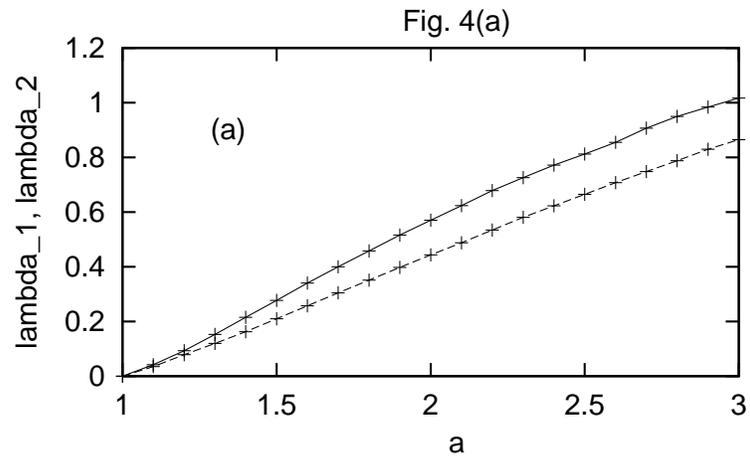
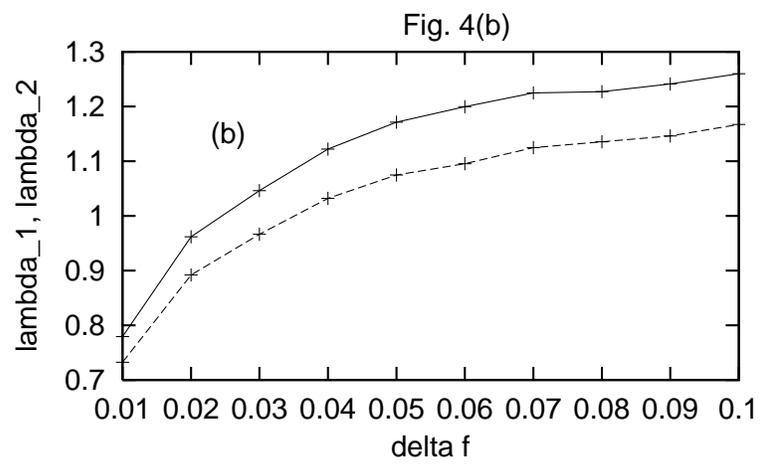
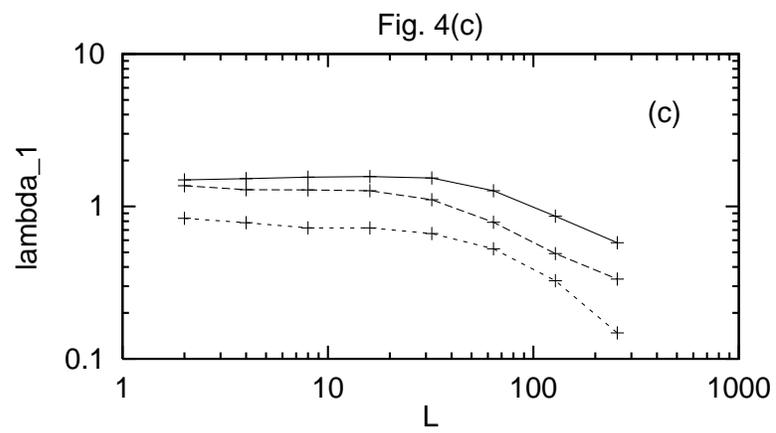
\end{document}